\documentclass[final,3p,times,twocolumn]{elsarticle}
\usepackage{graphicx}% Include figure files
\usepackage{dcolumn}% Align table columns on decimal point
\usepackage{bm}% bold math
\usepackage{psfrag}
\usepackage{amssymb}
\usepackage{amsthm}
\usepackage{dsfont}

\begin{document}
\begin{frontmatter}
\title{Diagonal stripes in the spin glass phase of cuprates}
\author[btu]{G. Seibold}
\address[btu]{Institut f\"ur Physik, BTU Cottbus, PBox 101344,
         03013 Cottbus, Germany}
\author[rome]{J. Lorenzana}
\address[rome]{SMC-INFM-CNR and
Dipartimento di Fisica, Universit\`a di Roma ``La Sapienza'', P.le Aldo
Moro 5, I-00185 Roma, Italy}

\begin{abstract}
Based on the unrestricted Gutzwiller approximation
we study the possibility that the diagonal incommensurate spin scattering
in the spin glass phase of lanthanum cuprates originates from stripe formation.
Similar to the metallic phase two types of diagonal stripe structures
appear to be stable: (a) site-centered textures which have one hole per
site along the stripe and (b) ferromagnetic stair-case structures which
are the diagonal equivalent to bond-centered stripes in the metallic phase
and which on average have a filling of $3/4$ holes per stripe site.  
We give a detailed analysis of the stability of both diagonal textures
with regard to the vertical ones.
\end{abstract}


\end{frontmatter}

In lanthanum cuprates (LSCO) the low energy incommensurate magnetic
scattering shows an interesting behavior upon entering the insulating phase, 
i.e. below doping $x_c\approx
0.055$.  According to the experimental findings 
\cite{waki99,waki00,mats00,fujita02}
the incommensurability $\delta$ (defined as the deviation of the 
magnetic peak from $Q_{AF}$) 
rotates from vertical above $x_c$ to
diagonal below $x_c$ where it is even detected in the 
elastic channel. Moreover, the orthorhombic lattice distortion allows one 
to conclude that
the diagonal magnetic scattering is one-dimensional with the associated
modulation along the orthorombic a-axis. Contrary, in the vertical phase 
the twinned LSCO samples show a two-dimensional scattering geometry
and the one-dimensionality of the stripe modulations is only suggested
by additional arguments. The magnitude of the incommensurability
is linear in doping $\delta=x$ also across the rotation but approaches
$\delta=x/2$ upon approaching the border of the AF phase at $x=0.02$.
Here we investigate the static properties of diagonal
stripe textures based on the Gutzwiller approximation (GA) of the one-band
Hubbard model. We use the following parameter set for the 
Hubbard model which fit the magnon dispersion
of undoped LCO and several other properties\cite{sei06}: On-site repulsion $U=7.5 eV$, Nearest-neighbor hopping
$t=340 meV$, Next-nearest neighbor hopping $t'=-0.2 t$. 

\begin{figure}[thb]
\includegraphics[width=7cm,clip=true]{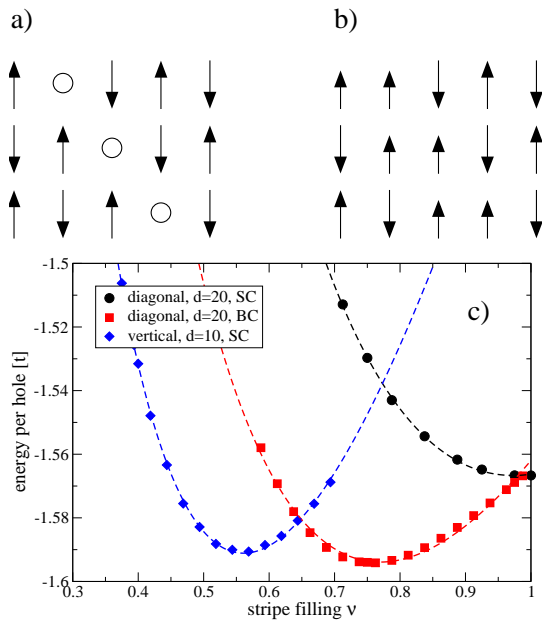}
\caption{Diagonal stripe textures investigated in this paper:
(a) site-centered stripe where non-magnetic sites are indicated by a circle
and (b) stair-case (bond-centered) stripes.
(c) Binding energy per hole as a function of stripe filling for
diagonal SC (circles), diagonal BC (squares) and vertical SC (diamonds)
stripes. The distance $d$ between stripes corresponds to the
separation of the domain walls along the $x-$ or $y-$ direction.
The dashed lines are fits as explained in the text.}
\label{fig12}
\end{figure}

Fig. \ref{fig12}a,b shows the two diagonal stripe textures which we found to be
stable saddle-points of our GA calculations. First we have the diagonal
site-centered (SC) texture (Fig. \ref{fig12}a) where the AF order-parameter
changes sign on the stripe sites similar to the corresponding
vertical structure in the metallic phase. Second we find a stair-case  
structure (Fig. \ref{fig12}b)with the sign change occuring on the bonds.
Contrary to vertical bond-centered textures (BC) the diagonal topology
implies a net ferromagnetic moment of the stripes. Note that diagonal
BC (DBC) stripes which are separated by an even number of unit cells along
$x$ (or $y$) have opposite magnetization whereas for distances with an 
odd number of unit cells all stripes have the same moment so that the
layer becomes macroscopically ferromagnetic. 
Fig.~\ref{fig12}c reports the 
energy per added hole, $e_h=[E(N+N_h)-E(N)]/N_h$ vs. the stripe
filling $\nu=N_h/N_s$. 
Here $N_h$ denotes the number of doped holes,
$N$ is the number of sites and $N_s$ the number of domain walls. 
For comparison we also show the corresponding curve for a vertical SC texture
(the vertical BC is identical).
Note that the curves in Fig. \ref{fig12}c correspond to low doping results 
in the sense that interaction effects between stripes are negligible.
From Fig. \ref{fig12}c it turns out that the preferred filling of 
vertical stripes
($\nu_{opt}\approx 0.55$) increases to ($\nu_{opt}\approx 0.75$) for
DBC and  ($\nu_{opt}\approx 1$) for DSC stripes. In Ref. \cite{seib09}
we have shown that this can account for the doping dependent
incommensurability as observed experimentally 
\cite{waki99,waki00,mats00,fujita02}.

In order to analyze the differences between the different diagonal 
configurations and between vertical and diagonal stripes we follow the 
procedure detailed in
Ref. \cite{sei04} and expand $e_h$ in a power series in $\nu$:
\begin{equation}\label{eh}
e_h = A/\nu +B +C\nu .
\end{equation}
As can be seen from the dashed lines in Fig. \ref{fig12}c 
this relation provides an
excellent fit to the data points for both vertical and diagonal stripes.
The coefficients in the expansion can be interpreted as follows \cite{sei04}:
$A$ is the energy cost per stripe unit to create an antiphase domain wall in
the undoped (i.e. AF) system. The parameter $B$ corresponds to the chemical 
potential to add
one hole into the empty stripe, and C is related to the inverse local
compressibility incorporating kinetic and interaction energy
contributions but fixing the stripe separation.

For undoped SC stripes the chemical
potential is at the center of an active band 
(between lower and upper Hubbard band)
which width is $\sim z^2 W^0(t')$ ($\sim z^2 W^0(t)$)  
for diagonal (vertical) SC stripes. Here $W^0$ denotes the width of an 
uncorrelated quasi one-dimensional band and $z^2$ is the Gutzwiller hopping 
renormalization factor 
which for fixed $U$ behaves as $z^2=1-(\alpha/W^0)^2$. 
Instead, for undoped BC diagonal stripes the associated ferromagnetic 
stripe magnetization cancels Gutzwiller band narrowing for the active 
band. In this case the  chemical potential is
on the top of the band with width $W^0(t')$. 
From these considerations we can understand the difference in the fitting
parameters for the various textures which are reported in the table.

\begin{table}[htb]\label{tab1}
\begin{tabular}{|l|c|c|c|} \hline
& Vertical SC & Diagonal SC & Diagonal BC  \\ \hline\hline
A [t] & 0.29 & 0.53 & 0.36 \\ \hline
B [t] & -2.62 & -2.66 & -2.52 \\  \hline
C [t] & 0.92 & 0.56 & 0.60\\ \hline
\end{tabular}
\caption{Parameters (in units of $t$) obtained by fitting the $e_h(\nu)$ 
curves of $d=20$ SC and BC 
diagonal and $d=10$ vertical SC stripes to  the expression Eq. (\ref{eh}).
The resulting fits are shown in Fig. \ref{fig12}c.}
\end{table}

Creation of a vertical (diagonal) BC
stripe in an ordered AF requires the destruction of one (two) AF bonds
and the formation of one (two) ferromagnetic bonds per stripe unit
(diagonal exchange couplings contribute only a fraction 
$(t'/t)^2$ to the energy estimate and are neglected.)
In the N\'{e}el limit the associated energy is $A_{vert}= J/2$ and
$A_{diag}= J$, respectively, which is further reduced 
due to quantum fluctuations.
In case of SC stripes the argument becomes more subtle as explained in
Ref.\cite{sei04}. The essential point here is that the core behaves as a 
spin liquid which again leads to $A\sim J$. 

The parameter $C$ corresponds to the relaxation energy 
upon adding holes to the undoped system and is decomposed into
a kinetic and interaction contribution $C=C_K + C_I$. $C_K$ can be determined
by optimizing the variational parameters for small $\nu$ and then computing
$e_h$ {\it vs.} $\nu$ without further changing the variational
parameters, i.e. by solely filling the frozen bands. The resulting 
linear contribution in $\nu$ to $e_h$ gives the $C_K$ parameter, and since $C$ 
is known we also obtain $C_I$. From this analysis we find that the kinetic
contribution to $C$ for SC diagonal stripes is practically zero due to
the almost flat active band. For vertical SC and diagonal BC stripes the
kinetic contribution in both cases is $C_K \approx 0.2 t$.

The almost vanishing $C_K$ parameter makes the
diagonal textures more susceptible towards disorder, especially due
to pinning by impurity dopants which tend to localize the holes. In fact,
replacing $Cu$ by nonmagnetic $Zn$ ions \cite{matsuda06} 
shifts the incommensurability close
to the value predicted for SC diagonal stripes.

The cooperation of J.L. and G.S. is supported by the Vigoni foundation.

\end{document}